\documentclass[final,12pt,authoryear,times]{elsarticle}
\usepackage[utf8]{inputenc}
\usepackage{amsmath}
\usepackage{mathrsfs}
\usepackage{graphicx}
\usepackage{epstopdf}
\usepackage{enumitem}
\usepackage{siunitx}
\usepackage{float}
\usepackage{mathtools}
\usepackage{xcolor}
\usepackage{subcaption}
\usepackage{algorithm}
\usepackage{algorithmic}
\usepackage{textcomp} 
\usepackage{url}
\newcommand{\comment}[1]{} 

 \usepackage[modulo]{lineno}

\usepackage{amssymb}
\usepackage{graphicx}
\usepackage{float}
\usepackage[english]{babel}
\usepackage[normalem]{ulem}
\usepackage{amsmath}
\usepackage{amsthm}
\usepackage{amsfonts}
\usepackage{bm}

\usepackage{physics}
\usepackage{lineno}

\journal{Journal of the Mechanics and Physics of Solids}

\author[a]{Xuanhe Li}
\author[a]{Brendan Unikewicz}
\author[b,c]{S. Chockalingam}
\author[d]{Hudson Borja da Rocha}
\author[a,d]{Tal Cohen\corref{corresponding}}

\affiliation[a]{organization = {Department of Mechanical Engineering}, 
addressline = {Massachusetts Institute of Technology}, 
city = {Cambridge}, 
state = {MA},
postcode = {02139},
country = {USA}}

\affiliation[b]{organization = {Department of Aeronautics and Astronautics}, 
addressline = {Massachusetts Institute of Technology}, 
city = {Cambridge}, 
state = {MA},
postcode = {02139},
country = {USA}}

\affiliation[c]{organization = {Department of Mechanical Engineering}, 
addressline = {Purdue University}, 
city = {West Lafayette}, 
state = {IN},
postcode = {47907},
country = {USA}}

\affiliation[d]{organization = {Department of Civil and Environmental Engineering}, 
addressline = {Massachusetts Institute of Technology}, 
city = {Cambridge}, 
state = {MA}, 
postcode = {02139},
country = {USA}}

\cortext[corresponding]{Corresponding author. Email: Tal Cohen: talco@mit.edu,}

\begin{document}

\begin{frontmatter}



\title{Interfacial Cavitation with Surface Tension: New Insights into Failure of Particle Reinforced Polymers}


\begin{abstract}
Understanding and mitigating the failure of reinforced elastomers has been a long-standing challenge in many industrial applications. In an early attempt to shed light on the fundamental mechanisms of failure,  Gent and Park presented a systematic experimental study examining the field that develops near rigid beads that are embedded in the material and describe two distinct failure phenomena: cavitation that occurs near the bead in the bulk of the material, and debonding at the bead--rubber interface [Gent, A.N. and Park, B., 1984. Journal of Materials Science, 19, pp.1947-1956]. 
Although the interpretation of their results has not been challenged, several questions stemming from their work remain unresolved. Specifically, the reported dependence of the cavitation stress on the diameter of the bead and the counterintuitive relationship between the delamination threshold and the material stiffness. 
In this work, we revisit the work of Gent and Park and consider an alternative explanation of their observations, interfacial cavitation. A numerically validated semi-analytical model  shows that in {the} presence of surface tension, defects at the bead-rubber interface may be prone to cavitate at lower  pressures compared to bulk cavitation, and that surface tension can explain the reported length-scale effects. A phase-map portrays the distinct regions of `cavitation dominated' and `delamination dominated' failure and confirms that for the expected range of material properties of the rubbers used by Gent and Park, interfacial cavitation is a likely explanation. 
Crucially, this result offers a new avenue to tune and optimize the performance of reinforced polymers and other multi-material systems.   

\end{abstract}


\end{frontmatter}


\section{Introduction}
\noindent Failure of materials often initiates near the site of an imperfection. A void, an asperity, an inclusion, or an inhomogeneity, are all examples of such imperfections. Understanding how they emerge, how they induce stress localization and  failure,  and how to mitigate their effect,  has been a central and longstanding challenge for material scientists and mechanicians alike. In recent years, however,  with the expanding use of   highly  deformable  multi-material systems, this challenge  has become   a problem of pressing  technological need to enable  applications ranging from 3D printed organs \citep{ribeiro2017assessing}, deformable screens \citep{kim2023fabrication}, and batteries \citep{lu2022void}, to name a few.  

The type of imperfection,  the loading state, and the specific material system at hand, cooperate to determine the particular mode of failure  and thus the failure threshold. In highly deformable materials, such as ductile metals or rubber, a commonly observed mode of failure is cavitation: the spontaneous expansion of a void upon application of a finite pressure beyond a critical threshold --- the \textit{cavitation pressure}  \citep{gent1959internal, horgan1995cavitation}. If loading persists, the cavity expansion would ultimately lead to fracture \citep{rogers1960tensile,ashby1989flow,raayai2019intimate, kumar2021poker, kundu2009cavitation}. Hence, the cavitation pressure  is thought to determine the  load bearing capability of the material.  


Though cavitation in solids is typically considered to occur in the bulk, experiments in soft materials often report cavities that emerge near interfaces\footnote{Note that here we restrict our attention to failure that emerges from randomly distributed pre-existing defects. Hence, needle induced methods, such as \textit{Cavitation Rheology} \citep{zimberlin2007cavitation,crosby2011blowing} and \textit{Volume Controlled Cavity Expansion} \citep{raayai2019volume,chockalingam2021probing}, are not considered. }.  
 In rubbers,  two types of tests have  emerged most commonly: the first is known as \textit{the poker-chip experiment}  \citep{chiche2005cavity, guo2023crack, gent1959internal}, and the second employs beads that are embedded in the material to create a constraint as the material is put under tension \citep{gent1984failure,poulain2017damage}. We will refer to the latter as   \textit{the bead experiment}.

By now, the poker chip experiment is probably most associated with the investigation of cavitation in soft solids. In their seminal study, \cite{gent1959internal} applied tension to thin  circular samples and reported an \textit{``unusual rupture process''} whereby  failure occurred at a critical load that was independent of a material length scale and instead directly proportional to the shear modulus ($\mu$). To explain this finding they  developed an analytical model. The pressure needed to expand a spherical cavity embedded in an infinite medium was calculated considering  an incompressible neo-Hookean material behavior and it was found that beyond a critical pressure of $p_{bc}/\mu=5/2$ the cavity would grow indefinitely, thus explaining their results as a cavitation phenomenon. The conclusions of this work and the competition between cavitation and fracture have been extensively debated since \citep{kumar2021poker,raayai2019intimate,poulain2017damage,williams1965spherical}.  \cite{gent1969surface} later cautioned of a flaw in  earlier studies with the emergence of an apparent length-scale dependence and the potential role of surface tension that stabilizes small defects \citep{gent1990cavitation}. Notably Gent  concludes that \textit{``the fact that a theory appears to work does not mean that it is true''}. 

Despite the particular focus, in most studies, on cavitation that occurs in the bulk of the poker chip specimen, some authors have studied the emergence of cavities at the adhered interface \citep{chiche2005cavity,dollhofer2004surface}, and report a surprising result: the threshold for cavities to appear depends on the thickness of the sample. In absence of an analytical model for such interfacial failure, a bulk cavitation model was used to explain the observed trends and the apparent influence of surface tension. 

In contrast to the poker chip experiment, which was initially motivated as a way to study the bulk response of soft solids, the bead experiment was  conceived as a way to study the influence of fillers on the mechanical response of reinforced polymers \citep{oberth1965tear,gent1984failure}. To test the hypothesis that the strength of the bond between the filler particle and the polymer is primarily responsible for the enhanced tensile strength, \cite{oberth1965tear} embedded millimeter scale beads into transparent tensile samples and observed the failure that occurred. In all of their samples they report that  {the} initial failure occurred in the bulk, near the bead, but not on it. Hence, consistent with the cavitation model of \cite{gent1959internal}, the failure threshold was shown to be proportional to the elastic modulus, and thus independent of both the interface toughness and the length-scale. Several years later, \cite{gent1984failure} revisited this problem and extended the experimental investigation to account for a larger range of bead sizes and varying levels of bonding strength. However, contradicting the earlier study, their results show a clear dependence on the size of the bead. Surprised by this anomaly,  they suggest that it calls for further study (further details of this study are provided in the next section). The bead experiments were once again revisited some 30 years later \citep{poulain2017damage,lefevre2015cavitation,breedlove2024cavitation}. This time with modern experimental tools allowing a close look  {into} the process of nucleation and growth of cavities and cracks at spatial and temporal resolutions of $1\mu$m and $66.7$ms, respectively. Similar to the earlier studies, these authors conclude that \textit{``the initiation of damage occurs from defects ... that are sufficiently far away from the constraints''}. {However, while initial defects in the considered rubber materials are smaller than the optical limit (i.e. $<1$ $\mu$m) and may be as small as the pore size (i.e. $\sim 1$ nm) a gap  remains in understanding the process by which those initial defects grow to the micron scale. Moreover, existing studies do not offer an explanation for the anomalous length scale effect reported by  \cite{gent1984failure} and neglect the  influence of surface tension, which may be significant\footnote{In the next section we show that even moderate levels of surface tension can dominate the cavitation response and dictate the stability threshold, even for highly  strain-stiffening materials.  }.}

In this work, we will again revisit the work of \cite{gent1984failure} with the particular purpose of addressing the unresolved question of length-scale dependence. We will expand on the previous studies by: (i) considering the role of surface tension, and (ii) allowing for initial defects at the bonded interface.  We will show that when these two critical ingredients are included into the model, \textit{interfacial cavitation} may become the dominant precursor to failure. In doing so, we will take advantage of a recent study \citep{henzel2022interfacial}, which showed that the expansion of an interfacial defect reaches a critical limit similar to bulk cavitation at an applied pressure of $p_{ic}/\mu\sim7/2$. Notably, this threshold is also independent of length scale and is higher than the limit for bulk cavitation (i.e. $p_{ic}>p_{bc}$), implying that if loads are applied remotely, bulk cavitation will dominate, and thus supporting the assumption  in earlier studies. However, this result does not account for the potential roles of surface tension and delamination -- the last missing ingredients needed to determine the competition between bulk and interfacial failure. 

This manuscript is organized as follows: in the next section we will begin by summarizing the key observations of \cite{gent1984failure}, then in Section \ref{prel} we will recapitulate  formulae for bulk and interfacial cavitation, which serve as a foundation for this work. We define the problem setting and assumptions in Section \ref{set} and proceed to develop an analytical model for interfacial cavitation and delamination in Section \ref{analitical}. We compare our analytical model to numerical results obtained via Finite Element Analysis (FEA) describe in Section \ref{numerical}. Results and discussion are found in Section \ref{rnd} and concluding remarks are offered in Section \ref{conclusions}.

\section{The experimental observations of \cite{gent1984failure}}
\noindent In their experiments, \cite{gent1984failure} performed tests using tensile specimens made of various transparent  elastomers for which  {they measured the} elastic modulus in the range  $0.85\leq E=3\mu \leq3~$MPa and  {estimated} the fracture toughness  the range $500\leq G_c\leq5000$ J/m$^{2}$. Spherical glass beads were embedded along the axis of the specimens after applying a surface treatment that serves to modulate the degree of bonding. The bead  diameters varied in the range $60\leq D \leq 5000$ $\mu$m and it was ensured that the width of the sample was at least three times the bead diameter.  

\begin{figure}[h!]
    \centering
    \includegraphics[width=1\textwidth]{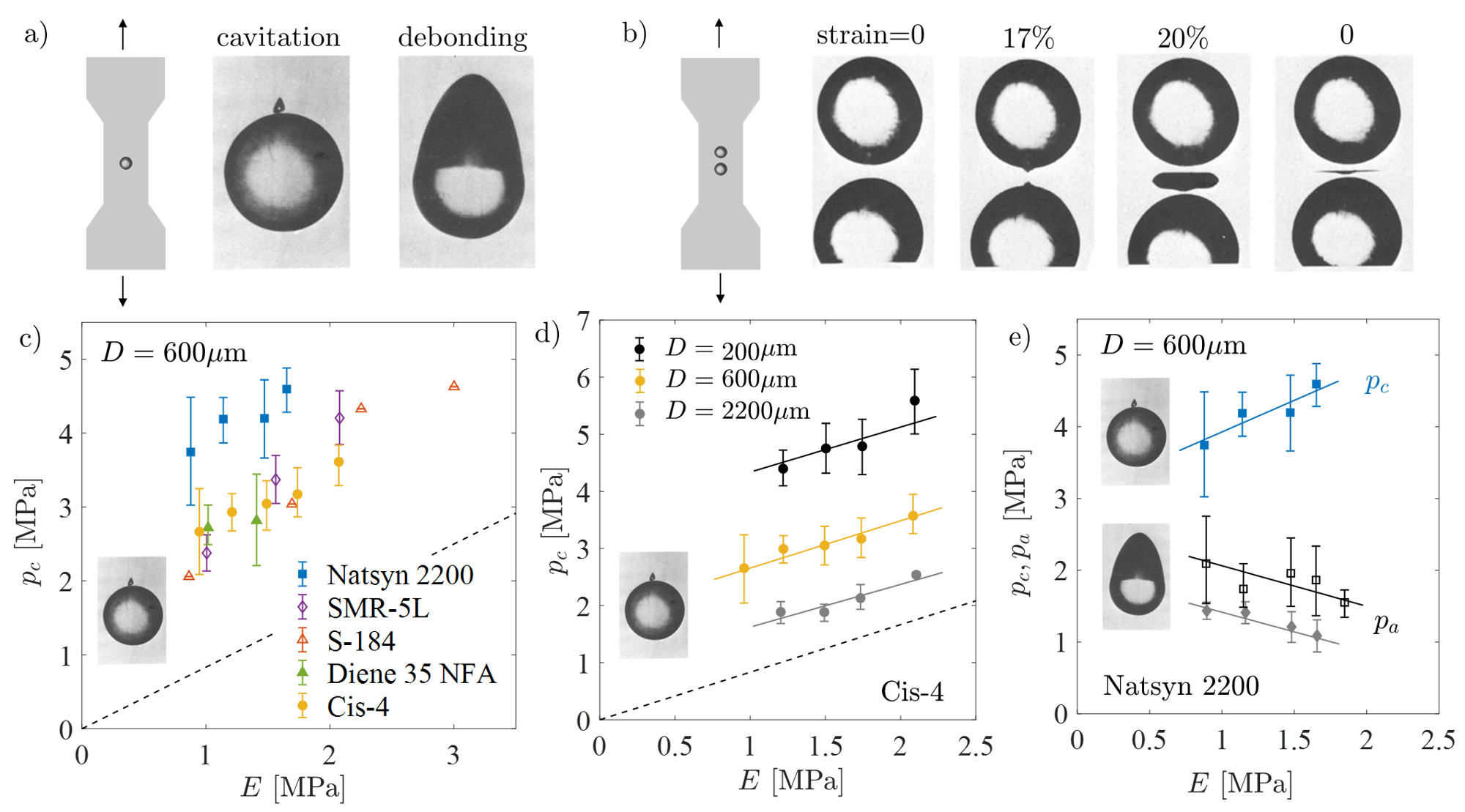}\vspace{-3mm}
    \caption{\textbf{Experimental results recreated from \cite{gent1984failure}. } (a) Illustration of loaded sample with embedded glass bead alongside typical observations of cavitation and debonding shown for  $1.2$ mm beads.  (b) Illustration of loaded sample with two embedded glass beads alongside a typical sequence of   cavitation progress shown for  $1.25$ mm beads in matrix of stiffness $E=2.2$ MPa. (c) Critical pressure at cavitation shown for different elastomers with stiffness modulated by varying degrees of cross-linking and with $600\mu$m diameter beads. (d) Influence of bead diameter on critical pressure at cavitation shown for poybutidiene (Cis-4) samples. (c,d) Dashed line denotes theoretical cavitation limit $(p_{bc}=5/6E)$. (e)  Influence of bonding strength on critical pressure shown for Natsyn 2200 with three different surface treatments (blue -- bonded, black -- untreated, grey -- treated).  {Results in (c-e) are shown for experiments with single beads as  in (a) and inserted images are provided to indicate the corresponding mode of failure.}
} 
    \label{gent_rep}
\end{figure}

Using linear elasticity and assuming the elastomers are incompressible, the authors estimated that the stress state at the poles of the bead is hydrostatic and double the applied load, i.e.   $p=2t_\infty$. It was verified in this work that for the specific experimental conditions this assumption holds well even when approaching the cavitation limit. 

Representative images of the two typically observed  behaviors are shown in Fig.\ \ref{gent_rep}(a). As described by the authors, for well-bonded particles a small cavity formed \textit{near} one of the poles at a critical applied stress, $p_c$, and then grew to eventually touch the glass as loading continued. With moderate bonding, abrupt delamination was observed at a critical pressure of $p_a$, after cavitation had already taken place. With weak bonding, abrupt delamination was observed and was not preceded by cavitation. When two well bonded beads were in the vicinity of each other (Fig.\ \ref{gent_rep}b), cavities were reported by the authors to first appear \textit{near} the poles of the beads then, as loading progressed, a larger cavity suddenly appeared midway between the two inclusions. 

The recorded critical pressures for well-bonded beads are shown in Fig.\ \ref{gent_rep}(c,d). For all of the materials considered, the critical pressure is shown to increase with the elastic stiffness. The authors note that although the measured values are above the theoretical cavitation pressure (dashed line), this trend agrees with the expected  theoretical trend for bulk cavitation. However, the bead diameter is shown to have a noticeable influence on the critical pressure (Fig.\ \ref{gent_rep}d); larger bead diameters correspond to lower critical pressure. In an attempt to explain this anomaly, \cite{gent1984failure} consider that the critical pressure results from tearing of a small crack close to the inclusion. Using the Griffith criterion with representative values of $E$ and $G_c$, they find a contradiction: the size of the initial  defect, $d$, would have to be as large as the bead itself (i.e. $d\sim D$). They note \textit{``Defects comparable to the size of the inclusion itself would certainly not escape notice in the experiments.''} 

Another unexpected result emerged upon modulating the bonding strength 
(Fig.\ \ref{gent_rep}e). While for well-bonded beads the critical pressure increases with the elastic modulus,  when the bonding is weak the trend is reversed. To explain this trend, \cite{gent1984failure} again attempted to use the Griffith criterion. However, it implies the opposite; that the critical pressure is proportional to both $E$ and $\Gamma$ --- \textit{the interfacial toughness}. Hence, they propose that stiffer materials (with higher $E$) might have lower $\Gamma$. They also note that the adhesive failure is delayed (i.e. higher $p_a$) for smaller beads, as in  cavitation.  

Overall, \cite{gent1984failure} conclude their manuscript with several open questions that emerged from their work, and have not been addressed to this day. In particular, the dependence of their results  on the bead size was not adequately explained as a fracture process using the Griffith's model. In this work, we will offer an alternative explanation.   

\section{Preliminaries}\label{prel}
\subsection*{Bulk Cavitation} 
\noindent Several studies have been devoted to theoretical investigation of the expansion of spherical cavities and the limit of cavitation that occurs in the bulk of the material  (see a review by \cite{horgan1995cavitation}). For the soft  elastomers considered here, the incompressible neo-Hookean material model is commonly employed, yielding the classical closed form analytical solutions of \cite{gent1959internal,gent1969nucleation}. Here we will recapitulate some of the key formulae that serve as a foundation for this work, using the present notation (see a useful review by \cite{dollhofer2004surface}).  

\begin{figure}[h]
    \includegraphics[width = 1.0\textwidth]{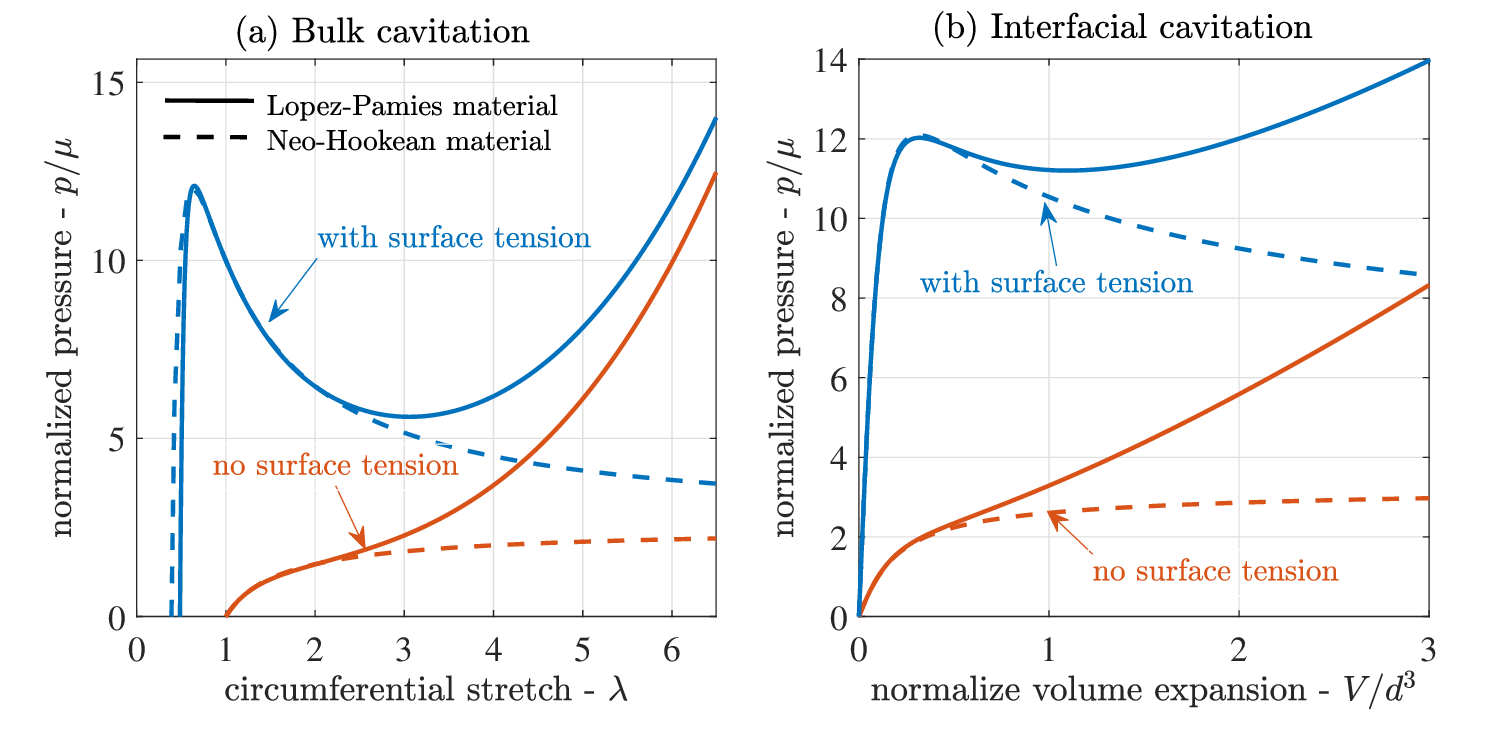}
   \caption{{\bf{Surface tension  {induces} cavitation instability in strain-stiffening materials.}} 
   Pressure-stretch response using neo-Hookean material  {model} and a two-parameter Lopez-Pamies material  {model} \citep{poulain2017damage} for PDMS with base to cross-linker ratio  of 20:1 with and without surface tension for (a) Bulk cavitation and (b) Interfacial cavitation. The pressure is normalized by the shear modulus $\mu$. Here for the case with surface tension we use the representative value of the normalized surface tension $\alpha = \gamma/(\mu d_0) = 2.5$. 
   }
   \label{fig:lopez}
\end{figure}

 Consider the expansion of a spherical defect, of initial diameter  $d_0$, embedded in an infinitely large medium. Let $\lambda_\theta$ denote the circumferential stretch and let $\lambda = d/d_0$ denote its value at the cavity wall (where $d$ is the expanded diameter), 
 the sum of applied internal pressure and remote tension,  $p$,  is then  related to the circumferential stretch at the cavity wall  by the following relation \citep{raayai2019volume}
\begin{equation}\label{eq:general_p}
    p(\lambda) =  \int_1^{\lambda} \frac{W^{\prime}(\lambda_\theta)}{1-\lambda_\theta^3} {\rm d}\lambda_\theta + \frac{4\gamma}{\lambda d_0},
\end{equation} where $\gamma$ denotes the surface tension, and $W(\lambda)$ is the strain energy density function specialized for the specific incompressible deformation field to be written as a function of the circumferential stretch. 
For the commonly used neo-Hookean material  $W(\lambda_\theta) =\frac{\mu}{2}(2\lambda_\theta^2 + \lambda_\theta^{-4}-3)$, with $\mu$ denoting the shear modulus, the above relation simplifies to  the explicit expression
\begin{equation}\label{eq:neo-p}
    p(\lambda)=\frac{\mu}{2}\left(5  - \frac{4}{\lambda}- \frac{1}{\lambda^4}\right) + \frac{4\gamma}{\lambda d_0}.
\end{equation}
In absence of surface tension ($\gamma=0$), the above formula for neo-Hookean material approaches an asymptotic limit as the cavity expands (i.e. $\lambda\to\infty$):  the classical bulk cavitation limit of $p_{bc}/\mu=5/2$. In presence of surface tension, a peak value emerges at a finite value of $\lambda$, as shown in Fig. \ref{fig:lopez}(a). In both scenarios, the presence of a maximum pressure indicates a critical threshold for spontaneous cavity expansion, namely a cavitation instability.

Elastomeric materials, as those used in the experiments of \cite{gent1984failure}, may exhibit strain-stiffening behavior. In absence of surface tension such materials are not expected to exhibit a cavitation instability \citep{breedlove2024cavitation}. Nonetheless,  it is instructive to examine the role of strain-stiffening on cavitation, in presence of surface tension\footnote{We would like to caution the reader of erroneous representations of the influence of surface tension on bulk cavitation in  the literature. The influence of $\gamma$, or more precisely its dimensionless counterpart -- $\gamma/(\mu d_0)$, is nonlinear and instability may occur for $\lambda<1$ due to the pre-existing stretch imposed  by surface tension (i.e. $p\neq 0$ when $\lambda =1$, as seen from \eqref{eq:neo-p}). }.  {To this end, we employ  the Lopez-Pamies material model \citep{poulain2017damage} and substitute it in \eqref{eq:general_p}. As seen from the curves with no surface tension (red) in  Fig. \ref{fig:lopez}(a), the  Lopez-Pamies and the neo-Hookean models predict a similar response for moderate stretch levels, and diverge for $\lambda>2$. The Lopez-Pamies  model has been shown to capture the strain-stiffening response of PDMS and thus does not exhibit cavitation.   However, in presence of even moderate levels of surface tension (blue curves), a peak  pressure emerges before the models diverge and  gives rise to cavitation instability that is insensitive to the specific constitutive model.} Hence, surface tension plays a dominant role in determining the threshold for onset of instability. A similar observation is made for the case of interfacial cavitation, as show in  Fig. \ref{fig:lopez}(b). The details of the approach used to generate these plots will be introduced in the following sections. We will then use the results 
to compare the peak value of the bulk cavitation to predictions of interfacial cavitation.  
Notably, since in both cases the particular choice of constitutive model does not significantly affect the qualitative behavior at onset of cavitation, we will restrict our attention to the neo-Hookean model from here on.

\subsection*{Interfacial Cavitation} 
\noindent In analogy to bulk cavitation, \cite{henzel2022interfacial} studied the expansion of cavities that initiate from the interface between the soft material and a rigid substrate, accounting also for delamination,  and  validated their predictions using data from \textit{pressure induced fracture} experiments  \citep{wahdat2022pressurized}.  They showed that the total energy invested in expanding an interfacial cavity can be decoupled into its elastic $(E_e)$ and delamination $(E_d)$ contributions, where 
\begin{equation}\label{eq0:Et}
    E_t(V,d)=E_e(V/d^3,d)+E_d(d),
\end{equation} with
\begin{equation}\label{eq:g0_Es}
    E_e=\int \displaylimits_0^V p{\rm d}V=\mu d^3 f(V/d^3),\qquad \text{and} \qquad E_d=\Gamma\frac{\pi}{4}(d^2-d_0^2), 
\end{equation}
such that the elastic energy is independent of the initial defect size $(d_0)$, and the energy invested in delamination is independent of the cavity volume $(V)$. Accordingly, the applied pressure can be written in its dimensionless form as
\begin{equation}\label{eq:g0_p}
\frac{p}{\mu}=f'(V/d^3).
\end{equation}
The function $f$ was obtained numerically for neo-Hookean material\footnote{{The function $f$ is provided in \cite{henzel2022interfacial} and included in the supplementary} material for completeness.} and it  was shown  that in absence of delamination $f'$ approaches an asymptotic cavitation limit -- $p_{ic}/\mu\sim 7/2$ (Fig.\ \ref{fig:PS}).
Then, an energy minimization procedure was used to determine also the delamination process and its influence on the critical pressure. 
However, once surface tension is considered, the above decoupling of the energy contributions is no longer possible. The surface tension influences the shape of the cavity, which influences the elastic energy and thus the applied pressure. In the following sections we will describe both a numerical and an analytical approach to overcome this complexity and to shed light on the role of surface tension in interfacial failure.

\section{Problem setting and assumptions}\label{set}
\noindent To capture the effect of surface tension on the critical threshold for onset of  interfacial cavitation, we choose the same problem setting as in  \cite{henzel2022interfacial}. Namely,  we consider a semi-infinite body whose surface is bonded to a rigid substrate everywhere except for  a circular region of diameter $d$ --  \textit{the defect} -- where there is no bonding. We define a Cartesian coordinate system $(X,Y,Z)$ with its origin at the center of the defect, such that the undeformed body occupies the region 
\begin{equation}
X,Y \in (-\infty,\infty), \qquad Z\in[0,\infty),
\end{equation} 
and the defect occupies the interfacial surface region $\mathcal{D}$ defined by 
\begin{equation} \label{eq:defect}
    X^2+Y^2\leq d^2/4, \qquad Z=0.
\end{equation}

\begin{figure}[h]
    \centering
    \includegraphics[width=0.6\textwidth]{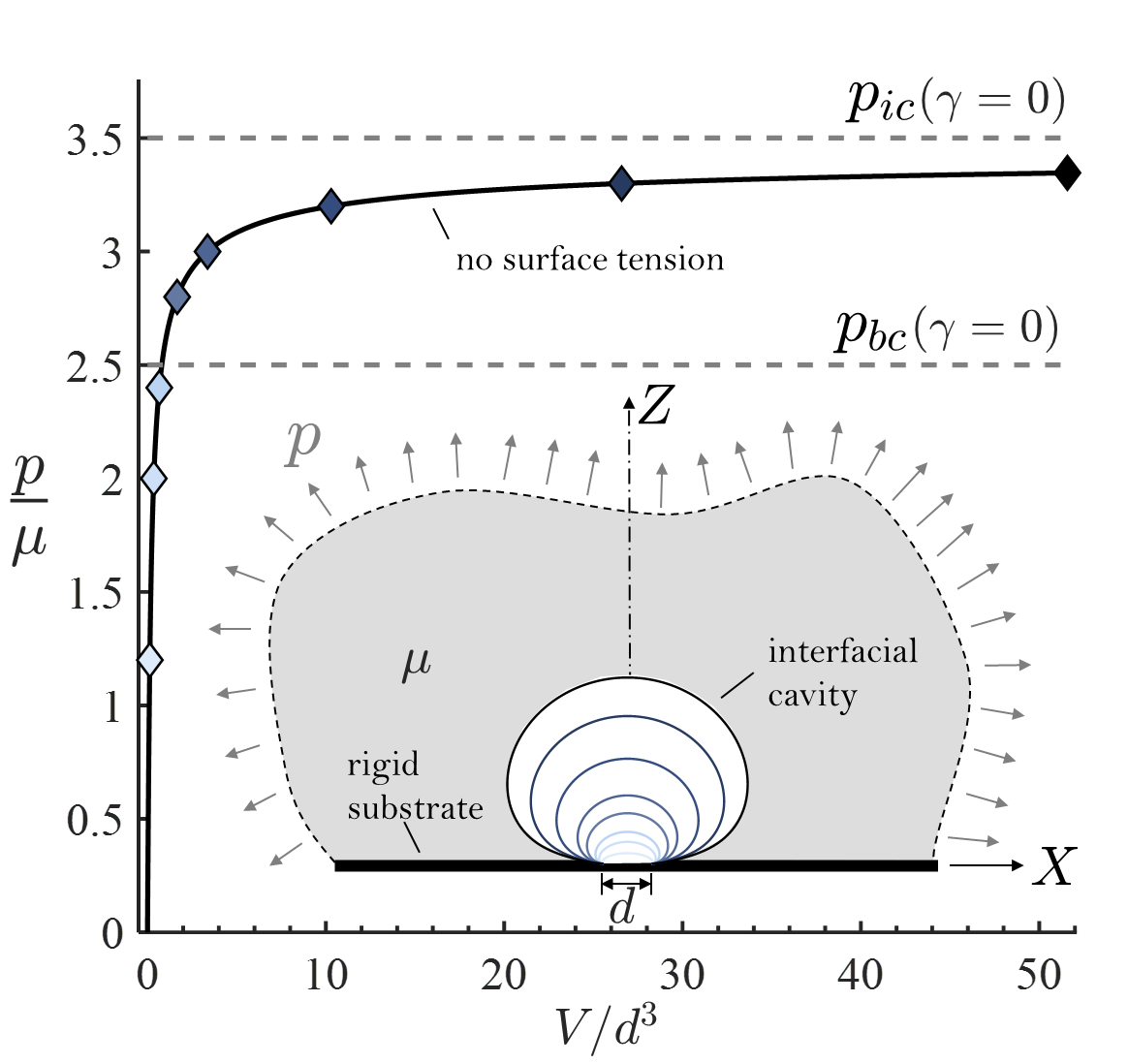}\vspace{-3mm}
    \caption{\textbf{Problem setting and results in absence of surface tension.} The black curve shows the relationship between normalized cavity volume and applied pressure for case without surface pressure, reproduced from \citep{henzel2022interfacial}. Grey dashed lines represent the bulk and interfacial cavitation pressures $(p_{bc}, p_{ic})$ without surface tension $(\gamma=0)$, respectively.   Illustration shows shapes of the interfacial cavity is at grows (in absence of surface tension) where curve shades correspond to the diamond markers along the pressure-volume curve.   The  pressure  can be applied remotely (as illustrated), or internally at the cavity wall.    }
    \label{fig:PS}
\end{figure}

Upon loading, either by application of remote tension or  internal pressure, the defect expands. We assume that throughout the expansion process, the volume of the defect is controlled and the axial symmetry, about the $Z$--coordinate, is preserved (Fig. \ref{fig:PS}). 

To compare our results with the classical relationship for bulk cavitation in presence of surface tension \eqref{eq:neo-p}, we restrict our attention to materials that are well modeled as incompressible  neo-Hookean solids. 
An important consequence of incompressibility is that the deformation field, and thus the critical threshold for cavitation, are identical if the load is applied as a remote tension, or as an internal pressure \citep{henzel2022reciprocal}; both of which have practical significance.  The former, is more directly relevant to this work and  corresponds to the load bearing capabilities of materials that may be prone to  failure from randomly distributed pre-existing defects within them, while the latter corresponds to internal growth, as seen for example due to formation of biofilm  \citep{borja2024onwards,nijjer2023biofilms}, is essential for evaluation of adhesion properties  \citep{wahdat2022pressurized,yamaguchi2018simple}, and can emerge due to phase separation \citep{kothari2023crucial}. 

We note that while the study by \cite{gent1984failure} considers cavities that form near rigid spherical inclusions and even identifies a clear influence of the inclusion size on the experimental  results  {for experiments with single beads}, in this study we limit our analysis to flat substrates. As such, we restrict our analysis to  defects that are small compared to the bead  {and to the dimensions of the specimen}, so that the local curvature  {and the layer thickness, do} not influence the cavitation threshold\footnote{ {Note that finite thickness of the sample, or the presence of a second bead at a finite distance, can influence the stress distribution and thus the failure response and its location \citep{guo2023crack,poulain2017damage}.}}. We can write the former restriction as   $d/D\ll 1$, where $D$ represents the diameter of the bead. With initial defect sizes below the optical limit, i.e. $d<1\mu$m, and reported  bead sizes in the range $60 \mu$m$\leq D\leq5000\mu$m, for the work of \cite{gent1984failure}, we have  $d/D \sim 10^{-3}$, consistent with our assumption. In the next sections, we will further examine the consequence of this assumption and will provide an alternative explanation for the apparent role of the bead size. Most importantly, this allows us to focus on the local stability criteria. Whereas, if the stress field is non-unifurm  prior to failure, the analysis becomes more complex.

\section{Analytical model}\label{analitical}
\noindent Considering the role of surface tension and preserving the volume $(V)$ and the defect size $(d)$ as our independent variables
the total energy in the system can be decomposed into three contributions
\begin{equation}\label{eq:g_Es}
    E_t(V,d)=\mu   F(V,d) +\gamma  S(V,d)+ \Gamma\frac{\pi}{4}(d^2-d_0^2).
\end{equation}
Here the first term corresponds to the elastic energy $(E_e)$, where we have introduced the  function $F$. The second term corresponds to the surface energy $(E_s)$, where $S$ represents the change in  surface area\footnote{This form of $E_s$ is consistent with the common definition of surface tension, whereby the change in surface energy is proportional to the change in surface area and hence  neglects any nonlinear (i.e. strain dependent)  effects \citep{bain2021surface,xu2017direct}.}.    The last term  corresponds to the delamination energy $(E_d)$, and remains unchanged from \eqref{eq:g0_Es}. 

Given the form of the total energy in \eqref{eq:g_Es} we can write the pressure as 
\begin{equation}\label{eq:pressure_def}
    p=\frac{\partial E_t}{\partial V}=\mu \frac{\partial F}{\partial V}+{\gamma}\frac{\partial S}{\partial V},  
\end{equation}
and, provided that the surface area is proportional to the defect area (i.e. $S\propto d^2$) two dimensionless numbers emerge from the above formulation, which reflect the competition between bulk and surface energies 
\begin{equation}\label{eq:defs}
   \alpha=\frac{\gamma}{\mu d_0}, \qquad \frac{1}{\varphi}=\frac{\Gamma}{\mu d_0}.
\end{equation}


Prior to making any simplifying assumptions, both functions $F$ and $S$ are unknown and their form is expected to depend on the specific set of material parameters $(\mu,\gamma)$. Obtaining $F$ and $S$ numerically, as in \cite{henzel2022interfacial}, is possible for a given value of $\alpha$ (see next section), but would be prohibitive over a continuous range, which is needed to evaluate cases of combined surface tension and interfacial failure (i.e. with finite $\Gamma$). Hence, in this work we make two simplifying assumptions:
\begin{enumerate}
    \item We assume that given the local nature of  surface tension, its influence  on the elastic  deformation field,  which spans indefinitely, and thus  on the total elastic energy for a given expansion volume (i.e. $V/d^3$) is negligible. Hence, we employ the known function for $\gamma=0$, such that
    \begin{equation}
      F(V,d) = d^3 f (V/d^3).
    \end{equation} 
    \item We propose two simplified alternatives for the function $S$. First, we calculate the change in surface area  from  the numerically derived shapes (as illustrated in Fig. \ref{fig:PS}) obtained without surface tension $(\gamma=0)$ -- we denote this set as\footnote{See corresponding function of $s_n$ in the supplementary material.} $S_n=d^2 s_n(V/d^3)$. 
    Second, we will simplify the shapes of the interfacial cavities to spherical caps, which are explicitly defined for given values of $V$ and $d$ (see formula in \ref{app:sphere}) -- we denote this  set as $S_O=d^2 s_O (V/d^3)$.
    Notably, we expect the former simplification  to be accurate when surface tension effects are small (i.e. $\alpha\ll1$) and the latter  when surface tension is dominant  (i.e. $\alpha\gg1$) thus serving as theoretical bounds on the expected result. 
\end{enumerate}
The above assumptions provide us with approximate functions $F$ and $S$, thus completing the analytical representation of the interfacial cavity expansion problem, accounting for both surface tension and delamination. To evaluate the consequence of the above assumptions we will compare results of the analytical model with numerical simulations in the next section. First, we will describe the solution procedure used to obtain the analytical results. 

\subsection*{Solution Procedure}
\noindent For a prescribed volume $(V)$, the total energy invested in the cavity expansion process \eqref{eq:g_Es} depends on a single free variable, the current size of the defect -- $d$.   In  equilibrium, the configurational forces in the system must vanish and thus the defect size  would stabilize at a value that minimizes the total energy in the system. Accordingly, provided the approximate forms of the functions $f\equiv F/d^3$ and $s=S/d^2$, along with the  definitions in \eqref{eq:defs}, we write the dimensionless energy as 
\begin{equation}\label{eq:g_Es_dl}
    \mathcal{E}=\frac{E_t}{\mu d_0^3}=\frac{d^3}{d^3_0} f\left({V}/{d^3}\right)+\alpha \frac{d^2}{d^2_0}  s\left({V}/{d^3}\right) + \frac{\pi}{4\varphi}\left(\frac{d^2}{d^2_0}-1\right).
\end{equation}
The solution procedure seeks to find a defect size, $d$, that minimizes this energy. This can be formally written  as
\begin{equation}\label{eq:minimization}
    d=\arg \min_d \left[\mathcal{E}(d; V, d_0,\alpha,\varphi) \right],
\end{equation}
or equivalently by requiring the derivative to vanish, such that 
${\partial\mathcal{E}}/{\partial d}=0$.
Provided the equilibrium value of $d=d(V)$ from this procedure, the corresponding pressure can be directly calculated from \eqref{eq:pressure_def} to write
\begin{equation}\label{eq:pressure_calc}
    \frac{p}{\mu}=\frac{\partial E_t}{\partial V}=f'(V/d^3)+\alpha \frac{d_0}{d} s'(V/d^3). 
\end{equation}

\subsection*{Stability Threshold}
\noindent In absence of both  surface tension and delamination, the cavitation instability   manifests as an asymptotic limit of the applied  pressure  with increasing expansion volumes. This limit can be expressed in terms of the second derivative of the energy function, as 
\begin{equation}
 \lim_{V\to\infty} \frac{\partial^2 E_t}{\partial V^2}= \lim_{V\to\infty}\frac{\partial p}{\partial V}=0,
\end{equation} which is seen also from the bulk cavitation result \eqref{eq:neo-p} and from Fig. \ref{fig:PS}. 

In presence of surface energies and thus with the introduction of length-scale dependence,  the instability can emerge at finite expansions.  Considering surface tension, but without delamination (i.e. with $d\equiv d_0$), this stability threshold can be written as 
\begin{equation}
   V_\gamma= \arg \left[ \frac{\partial^2 E_t}{\partial V^2}=0\right]= \arg \left[ \frac{\partial P }{\partial V}=0\right],
\end{equation} where $V_\gamma$ denotes the corresponding critical volume. 

Now, if we consider both surface tension and delamination, a competition between the two instability modes emerges. Delamination would initiate at the first expansion for which a solution is found in \eqref{eq:minimization}, which can be written as 
\begin{equation}
    V_\Gamma=\arg \left[\frac{\partial\mathcal{E}}{\partial d}(d_0; d_0,V,\gamma,\varphi)=0\right],
\end{equation}
 where $V_\Gamma$ denotes the corresponding critical volume. 

By comparing the above two stability thresholds we can determine the critical threshold for the system, namely
\begin{equation}\label{Vc}
V_c=\min[V_\gamma,V_\Gamma],
\end{equation}
 and correspondingly the critical pressure, $p_c=p(V_c/d_0^3)$, is calculated from \eqref{eq:pressure_calc}.


\section{Numerical Simulation}\label{numerical}

\noindent Numerical simulation of interfacial cavity expansion, without delamination ($d \equiv d_0$), was implemented using the open-source finite element python package FEniCS \citep{AlnaesEtal2015} with 2D-triangular elements.  The finite element code has been made available online\footnote{\url{https://github.com/cohen-mechanics-group/interfacial-cavitation-surface-tension}}. In this section we will first summarize the key aspects of our finite element simulation, including the implementation of surface tension. Then, we will use the simulation results to verify   the  simplifying assumptions we made in the development of the analytical model in the   previous section.


\subsection*{Problem setting}
\noindent The theoretical model assumes that the interfacial cavity is embedded in a  semi-infinite body, with its surface bonded to a rigid substrate except for a circular region of diameter $d$. To approximate this setting in the numerical simulation we consider a cylindrical domain of diameter $W$ and height $H$, where we ensure both $W$ and $H$ are much larger than $d$ to eliminate  boundary effects. Specifically, in the simulation we choose $W/d = 100$ and $H/d = 50$. 

The bottom surface of the cylinder is fixed everywhere except for the central circular region of diameter $d$, which is traction-free. The top surface is subjected to a uniform normal displacement, and a roller boundary condition is applied to the lateral surface to ensure that the remote region is in a hydrostatic stress state. Given the geometry and the boundary conditions of the problem, we make use of the axial symmetry of the system and thus reduce the simulation to a 2D representation on one cross-section. For the constitutive relation, we employ the neo-Hookean model and adopt the 
$(\bm{u},p)$ approach \citep{stewart2023magneto} to account for near-incompressibility, where the displacement $\bm{u}$ and the pressure-like field $p$ are treated as separate degrees of freedom.

\subsection*{Implementation of surface tension}
\noindent A method for numerical  implementation of surface tension coupled with elastic deformation (i.e. elasto-capillarity) within a finite element framework has been proposed  by \cite{henann2014modeling}. Therein surface tension is  directly imposed as a traction force which is prescribed as a function of the local mean curvature. Here, however, leveraging the capability of FEniCS to conduct automatic differentiation \citep{logg2012automated}, we implement the surface tension by introducing its energetic contribution,  as shown in \eqref{eq:g_Es}, which can be recast in the integral form using Nanson's formula
\begin{equation} \label{eq:num_surf}
    E_s = \gamma S = \gamma\int_\mathcal{D}  ||\bm{F}^{-T}\bm{n_R}||\ d A,
\end{equation}
such that the surface area of the deformed cavity is obtained by integration of the deformation field, where $\bm{F}$ is the deformation gradient tensor,  $\bm{n}_R$ is the unit normal vector int he reference configuration, and  $dA$ is the reference  unit area of the defect. 

With the surface energy incorporated into the total energy, we solve the nonlinear variational problem directly in FEniCS. 
As a validation,  we compared this approach to the one proposed by \cite{henann2014modeling} and good agreement has been found between the results obtained from these two approaches.



\section{Results and Discussion}\label{rnd}

\noindent In this section, we will first discuss the theoretical results for interfacial cavitation without delamination obtained from \eqref{eq:minimization} and \eqref{eq:pressure_calc} with $d\equiv d_0$. We will  compare the theoretical predictions with the experimental results of \cite{gent1984failure}, thus challenging the assumption therein  that the cavitation occurs in the bulk, from initial defects that are near the constraints but not on them \citep{poulain2017damage,lefevre2015cavitation}. Then, we will extend the analysis to examine the role of delamination and its ability to explain the experimental observations. The theory  will then be extended to account also for $d>d_0$ in \eqref{eq:minimization} and \eqref{eq:pressure_calc}.

\subsection*{Interfacial cavitation without delamination}

\begin{figure}[h]
    \centering
    \includegraphics[width=0.9\textwidth]{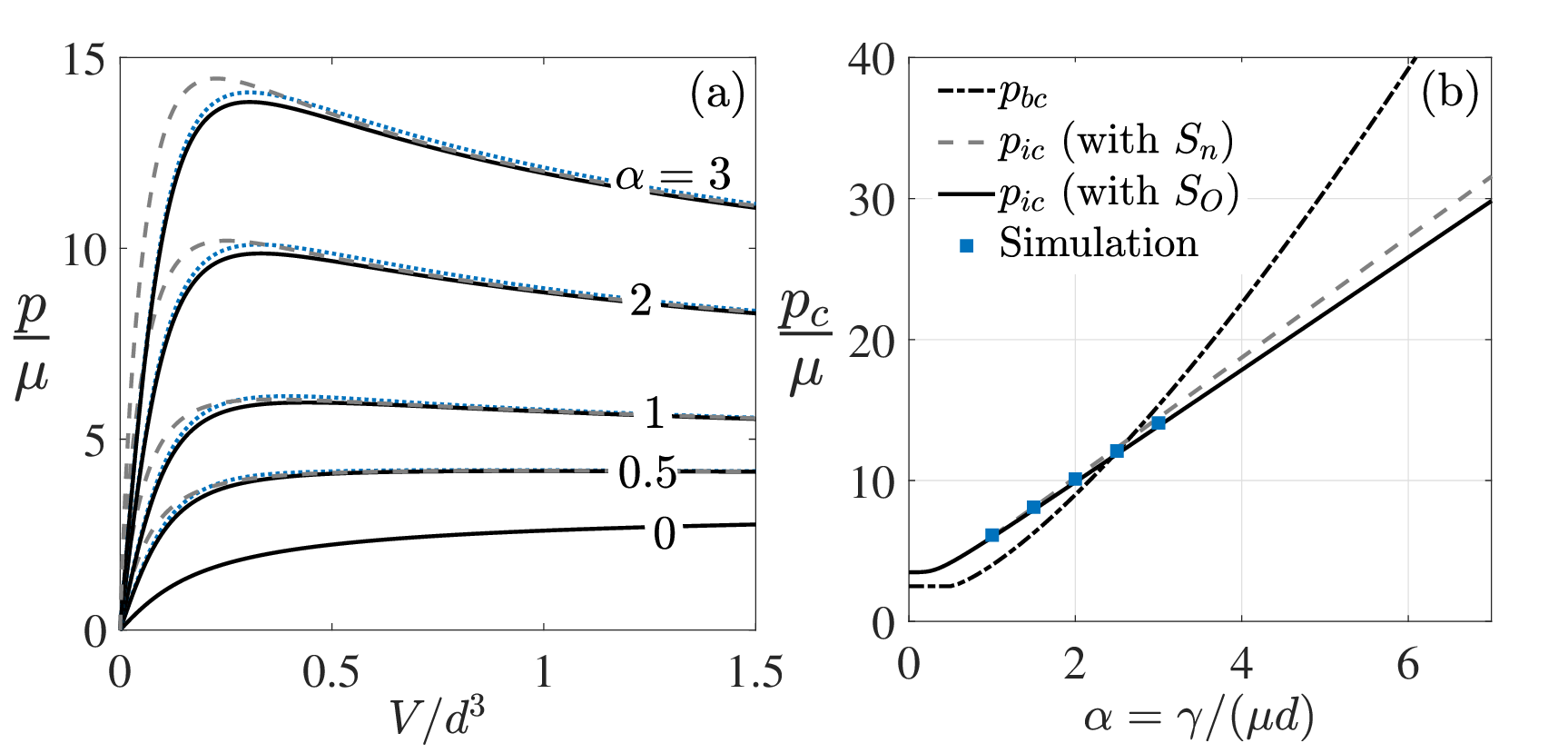}\vspace{-3mm}
    \caption{\textbf{The role of surface tension on interfacial cavitation, without delamination.} (a) Pressure volume response for varying levels of normalized surface tension $(\alpha)$ shown using three different models. Grey {dashed} curves correspond to analytical model using surface area approximation based on expansion without surface tension (i.e. with $S_n$); black curves correspond to the spherical cap approximation (i.e. with $S_0$); and blue {dots} correspond to the simulation results, in agreement with the legend on the right panel. (b) Critical pressures for interfacial cavitation correspond to the curves in (a). {Black dash-dotted curve corresponds to bulk cavitation.}  }
    \label{fig:SSC}
\end{figure}

\noindent The influence of surface tension on interfacial cavitation (without delamination) is shown in Fig.\ \ref{fig:SSC} for three different models: the analytical models described in Section \ref{analitical}, using approximate surface areas based on the expansion with no surface tension (with $S_n$) and using the spherical cap approximation (with $S_0$);  and the numerical simulation described in Section \ref{numerical} (shown in blue). From the pressure volume responses  in Fig.\ \ref{fig:SSC}(a) we find that  both analytical approximations  are in  good agreement with the numerically obtained results and become indistinguishable  as expansion progresses and elasticity dominates. At small expansions, where surface tension is dominant, the curves obtained using the spherical cap approximation (i.e. with  $S_O$)  agree well with the numerical curves. The comparison is seen also in  Fig.\ \ref{fig:SSC}(b) where the peak (critical) pressures obtained from the two theoretical models  closely agree with the simulation result.  

It is revealing to compare the critical pressures for surface and bulk cavitation. The dashed curve in Fig.\ \ref{fig:SSC}(b) shows the critical pressure for  cavitation in the bulk initiating from a defect with initial radius $d_0$, as obtained from \eqref{eq:neo-p}. Though at low levels of surface tension $\alpha \lesssim 2.5$, such a defect would be prone to exhibit instability  before interfacial cavitation ensues, for $\alpha \gtrsim 2.5$ interfacial cavitation dominates.  {Note however that this comparison equates between defects of similar initial diameters, at the interface or in the bulk although their geometry is different. If we compare, for example, cavities of similar initial surface area, then   interfacial cavitation would become even more favorable\footnote{This can be seen by rescaling $\alpha$ in \eqref{eq:neo-p}, so that $d=d_0$ corresponds to a sphere with effective surface area ($\pi d_0/4$), identical to that of the interfacial cavity.}. Moreover, it is not necessary that defect sizes would be similar in the bulk and at interfaces.} 

Note that while the bulk cavitation pressure curve in Fig.\ \ref{fig:SSC}(b) is nonlinear, the interfacial cavitation curves all approach a linear slope of $4$ for large values of  $\alpha$, where elastic contributions become negligible and the spherical cap approximation becomes accurate. This is consistent with the slope predicted by  \cite{kundu2009cavitation};  therein the spherical cap approximation was used to estimate the role of surface tension in needle based cavitation experiments.

\begin{figure}[h]
    \centering
    \includegraphics[width=0.75\textwidth]{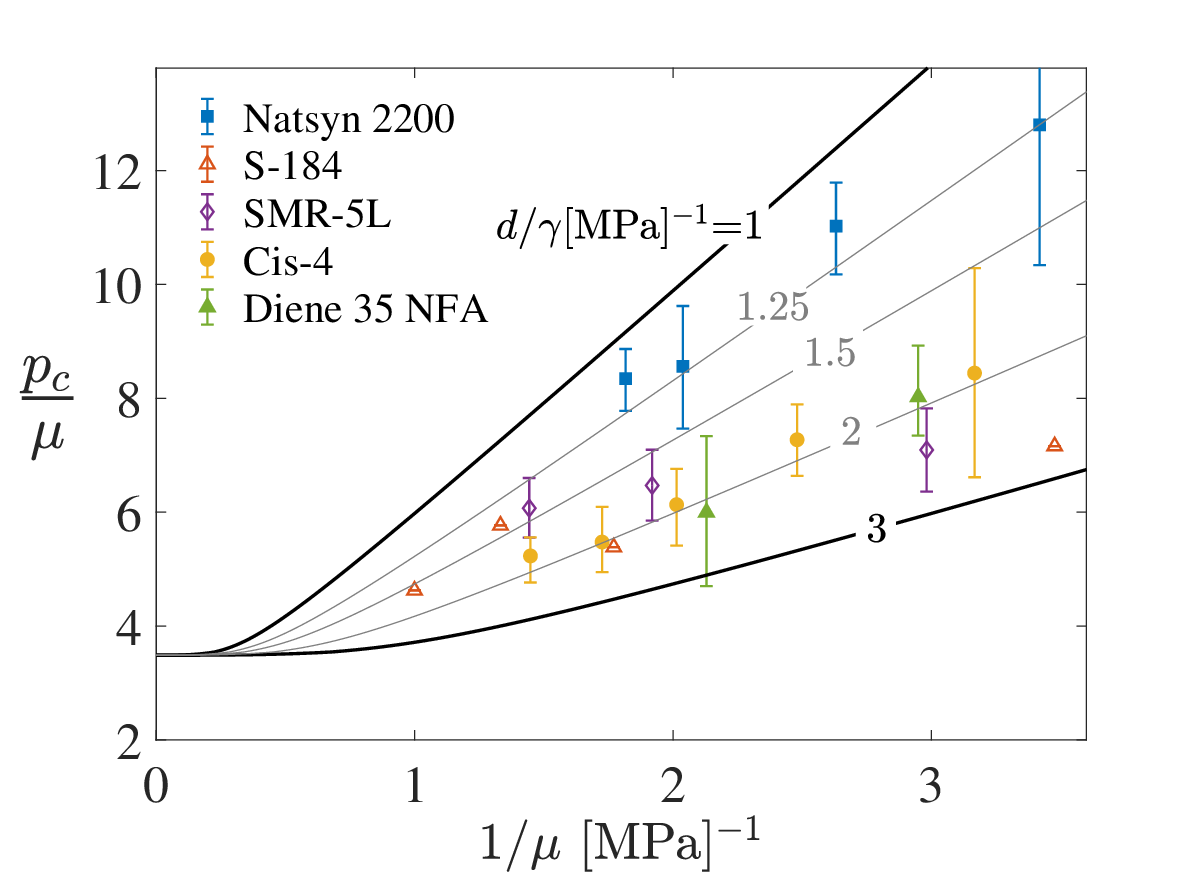}\vspace{-3mm}
    \caption{\textbf{Theoretical model  {explains} experimental trends  without delamination and for constant bead size.} Experimental results here are rescaled from Fig.\ \ref{gent_rep}(c) -- see caption therein for details. Theoretical curves are obtained via the analytical model detailed in Section \ref{analitical} with the  spherical cap approximation and shown for different  values of $d/\gamma$. }
    \label{fig:gent1}
\end{figure}

Due to the close agreement of the  critical pressure predictions of the three models shown in Fig.\ \ref{fig:SSC}(b), in all of the following analyses and comparisons, we will use the spherical cap approximation for its analytical simplicity. As a first step, we can examine the agreement of this model with the experimentally measured critical pressures of \cite{gent1984failure}, previously shown in Fig.\ \ref{gent_rep}.  {Though both surface tension and defect sizes can vary for different material systems and for different compositions, we find in Fig.\ \ref{fig:gent1} that most of the experimental data for the different materials seem to follow the theoretical  trend of constant $d/\gamma$. Results for  SMR-5L  can be seen as an outlier in this regard.} Moreover, all of the results are captured within a narrow range of $1\leq d/\gamma\leq 3$ [MPa]$^{-1}$, which corresponds to $1\leq\alpha\leq3.4$.  With surface tension for these materials estimated in the range $0.01\leq\gamma\leq 0.1$  [J/m$^2$] \citep{johnson1971surface}, a corresponding range of  defect sizes would be $0.01\leq d\leq 0.3$ [$\mu$m], which agrees with the expectation that the defects are initially smaller than the optical limit.   {This  could potentially  explain why nucleation in these cases is only observed once failure propagates away from the interface and into the bulk, and thus reported as a bulk cavitation phenomenon.}

\begin{figure}[h]
    \centering
    \includegraphics[width=0.7\textwidth]{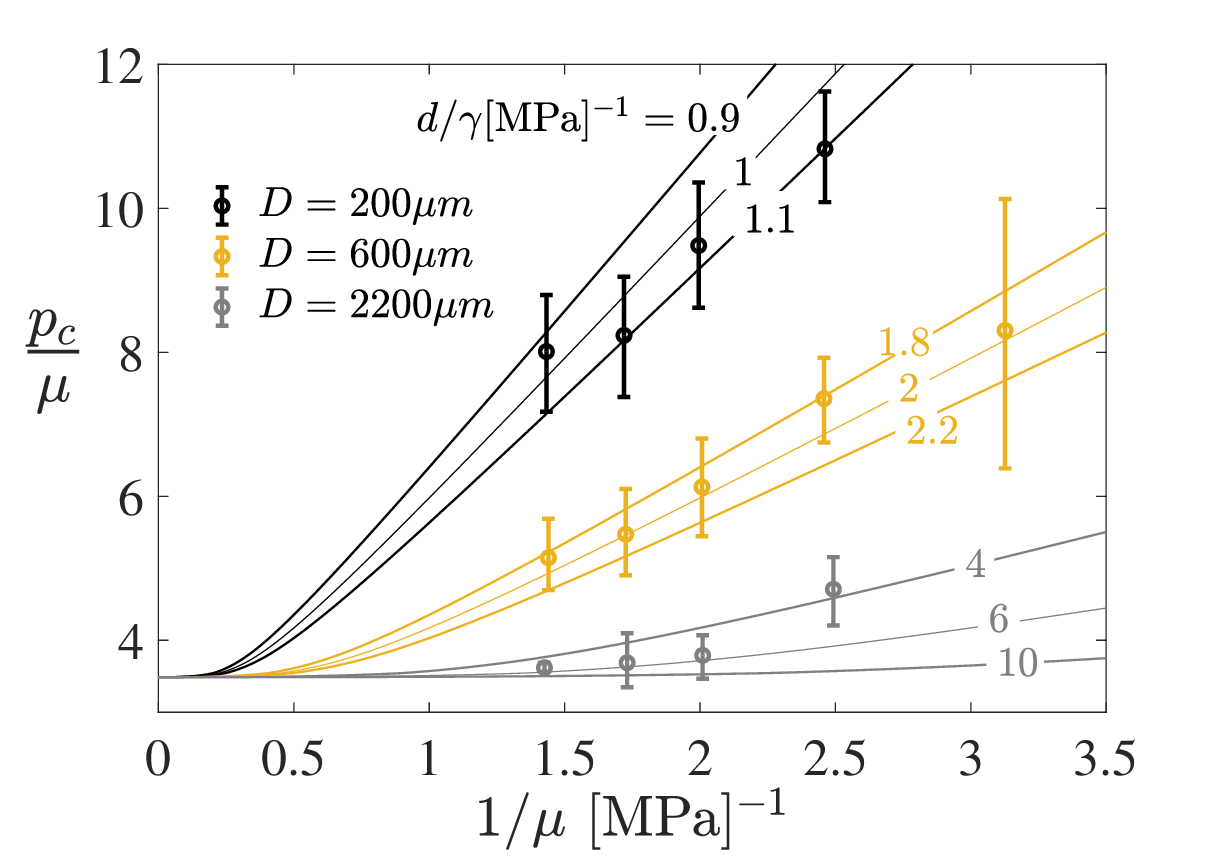}\vspace{-3mm}
    \caption{\textbf{Theoretical model  {explains} experimentally observed length scale dependence.}   Experimental results here are rescaled from Fig.\ \ref{gent_rep}(d) -- see caption therein for details. Theoretical curves are obtained via the analytical model detailed in Section \ref{analitical} with the  spherical cap approximation and shown for different  values of $d/\gamma$.   }
    \label{fig:gent2}
\end{figure}

Next, we examine the influence of bead diameter on the cavitation threshold by comparing the theoretical predictions with experimental data for Cis-4 (shown previously in Fig.\ \ref{gent_rep}d), as seen in Fig.\ \ref{fig:gent2}. We find that the theoretical model captures the reported  length scale effect, if the size of the largest interfacial defects (i.e. those that would fail first) is assumed to be related to the bead size.  These results support the hypothesis made by \cite{gent1984failure}, that the defect size can be considered proportional to the bead size (i.e. $d\propto D$). The same argument cannot be made for defects in the bulk.

\subsection*{The Role of Delamination}
\begin{figure}[h]
    \centering
    \includegraphics[width=1\textwidth]{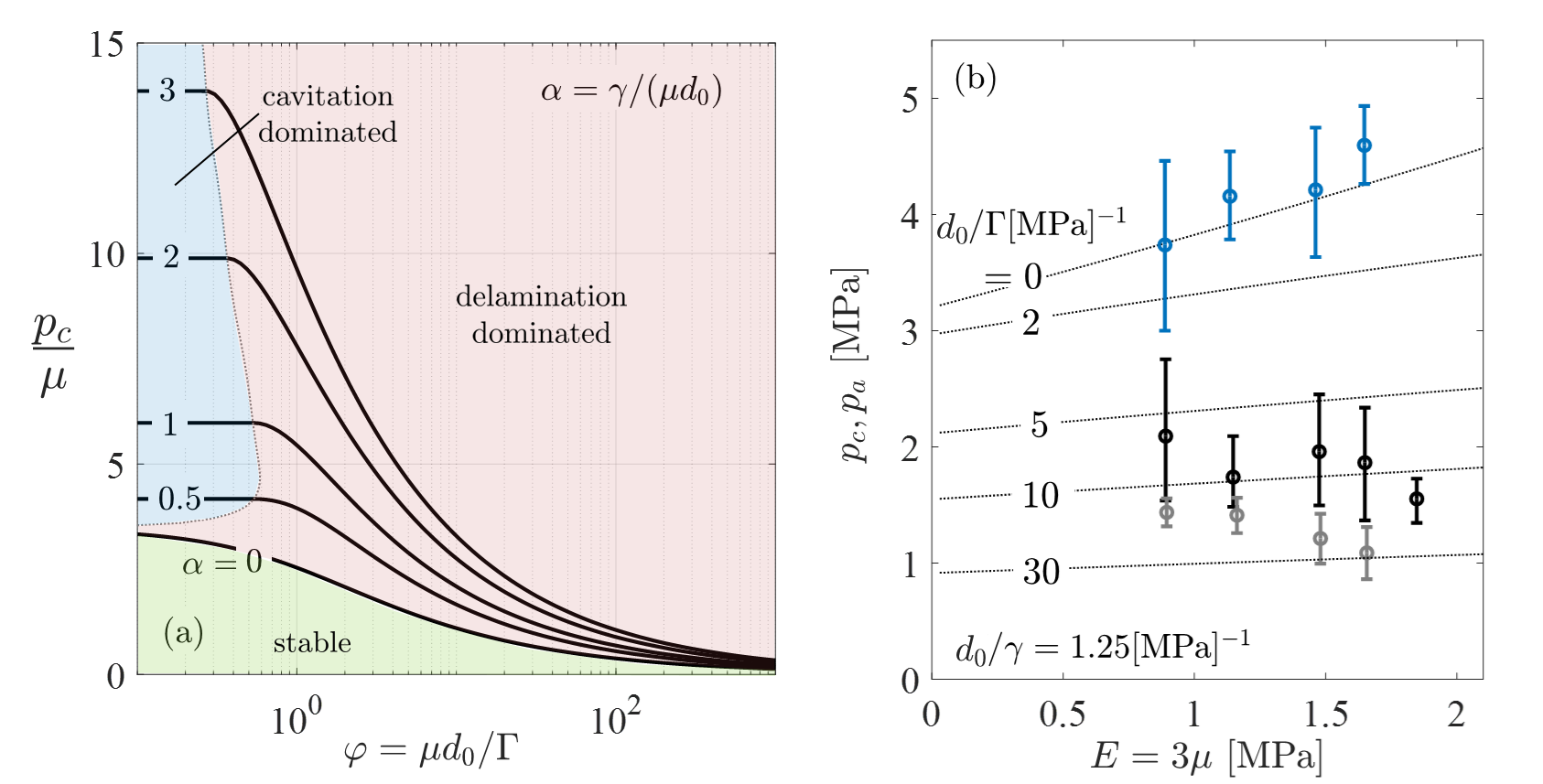}\vspace{-3mm}
    \caption{\textbf{Delamination can explain experimentally observed trends.} (a) Stability phase diagram: Black curves show theoretically predicted  critical pressure for varying levels of   $\varphi$ with different values of surface tension $\alpha$, as obtained via the analytical model detailed in Section \ref{analitical} with the  spherical cap approximation. Green shaded region corresponds to a stable regime. The blue region corresponds to cavitation instability dominated instability. The red region corresponds to delamination dominated instability. Notice that the horizontal coordinate is shown on logarithmic scale.  (b) Theoretical predictions of critical pressure for material with $d_0/\gamma=1.25$ [MPa]$^{-1}$ and for different values of $d_0/\Gamma$ are shown in comparison with experimental results  from Fig.\ \ref{gent_rep}(e) -- see caption therein for details.    }
    \label{fig:gent3}
\end{figure}

\noindent In absence of delamination, the results shown thus far correspond to a strictly bonded interface, namely to the limit $\Gamma\to\infty$, or equivalently to $\varphi\to 0$. In Fig.\ \ref{fig:gent3} we relax this constraint, allowing for $d>d_0$ in \eqref{eq:minimization} and \eqref{eq:pressure_calc}, to  examine the influence of finite interfacial toughness on the critical pressure. Fig.\ \ref{fig:gent3}(a) presents a phase diagram distinguishing between stable and unstable states of the material system. The green region corresponds to states in which no instability is expected for any value of $\alpha$ or $\varphi$;  it is bounded by the curve for $\alpha=0$, which is identical to the result obtained by \cite{henzel2022interfacial}.
  For finite values of $\alpha$, the critical pressure curves are shown to  initiate from the cavitation pressure at $\varphi=0$, which corresponds to values in  Fig.\ \ref{fig:SSC}(b), then a plateau is seen for moderate values of $\varphi$; in this region (shaded blue),  interfacial cavitation initiates before debonding becomes energetically favorable (i.e. $V_\gamma<V_\Gamma$ in \eqref{Vc}). For larger values of $\varphi$ (shaded red) the peak pressure is associated with the onset of delamination  (i.e. $V_\gamma>V_\Gamma$ in \eqref{Vc}) and is shown to decrease monotonically as $\varphi$ increases. At the limit of $\varphi\to\infty$, the curves unite as effects of both surface tension and elasticity are dwarfed by interfacial delamination,  which becomes the dominant and most energetically favorable  mode of cavity expansion. In this limit, the approximation of  linear elasticity  used by \cite{gent1984failure} applies and the normalized critical pressure scales as $1/\sqrt{\varphi}$. 

In Fig.\ \ref{fig:gent3}(b) we compare the theoretical prediction to the results of \cite{gent1984failure} previously shown in Fig.\ \ref{gent_rep}(e).  According to the comparison in Fig.\ \ref{fig:gent1}, the ratio $d_0/\gamma=1.25$ [MPa]$^{-1}$ serves as a good approximation for  Natsyn 2200; the material  used for the delamination experiments.  {Though the difference in surface treatment influences both surface tension and the defect size, we will use this value as a representative value to examine the qualitative trends observed in the experiments.}   Accordingly, we show dimensional curves of predicted critical pressure as a function of the elastic modulus for constant values of the ratio $d_0/\Gamma$. As expected, for the case of cavitation (blue data),  the theoretical curve with $d_0/\Gamma=0$ agrees well with the experiments and exhibits a distinctly positive slope. For increasing values of $d_0/\Gamma$ the theoretically predicted slope decreases.  {However, a negative slope is not reproduced for constant $d_0/\Gamma$ values. A negative trend might be explained by the interdependence between the interface toughness and the stiffness of the material. In particular, a reduction of interfacial toughness with increasing stiffness can lead to negative slopes.   Nonetheless,  by comparing the theoretical curves and the experimental data for weakly bonded beads we can estimate that for untreated beads (black data)  $d_0/\Gamma\sim 10$ and for treated beads (grey data) it can be approximated as  $d_0/\Gamma\sim 20$.  Accordingly, for these material we have $1<\varphi<10$, for which  nonlinear elastic effects are  appreciable.}  Finally, considering defect sizes  of the order of $d_0\sim 1$ [$\mu$m] for weakly bonded interfaces,  we find from this comparison that the interfacial toughness is  of the order $\Gamma\sim 0.1$ [N/m$^2$].  {However, we caution that though the qualitative response is captured, the interdependence between $\Gamma$, $\gamma$, and $d_0$, prohibits a  quantitative assessment of their values without obtaining additional independent measurements.}

\section{Conclusion}\label{conclusions}


\noindent Understanding the initiation of failure near rigid inclusions is essential to elucidate the toughening mechanisms in particle-reinforced polymers and to inform the design of various multi-material systems. 
The experimental work of \cite{gent1984failure} set-out to study such failure processes by examining the field that develops near rigid spherical inclusions embedded in transparent elastomers. Considering a range of bead sizes and material stiffnesses, their work identified two primary failure modes, depending on the strength of the bond, which were described as: (i) bulk cavitation near the beads (but not on them), and (ii) delamination. However, several questions remained open. Though cavitation  was expected to be independent of length-scale, their work found a dependence of the critical pressure on the size of the bead. In an attempt by to explain this effect using a fracture criteria  they found a contradiction; that the defect size would have to be unreasonably large. Another open question revolved around the critical threshold for delamination; while the cavitation threshold increases with increasing stiffness, the delamination threshold was found to  reduce with increasing stiffness. This trend, too, could not be explained using a delamination criteria. 

In this work, we  challenge the conventional understanding and offer  an alternative  explanation to the experimentally observed phenomena. 
 {Our analysis examines the possibility that the nucleation event, even if eventually seen in the bulk,  emerges from sub-optical defects at the interface. }  To study interfacial cavitation in these systems, our analysis forfeits the analytical convenience of spherically symmetric cavitation fields. Through the development of a numerically validated semi-analytical model, we find that for  levels of surface tension and defect sizes relevant to the considered material systems,  interfacial defects are prone to cavitate at a lower pressure compared to those in the bulk.  Furthermore, a phase diagram  clearly distinguishes two regimes of failure and, without any further assumptions, our analysis offers an explanation to the unresolved questions of \cite{gent1984failure}, capturing both the length-scale effect and the delamination  response. 
Overall, the results of this work reinforce the notion that interfacial properties govern the failure process and thus challenge the traditional explanation of bulk cavitation as a primary mechanism of failure in these systems.  {Importantly, our results are not in contradiction with the observation of micron-scale nucleation events in the bulk. A possible explanation is that following the nucleation from sub-micron defects at the interface,  subsequent growth of the nucleated cavities  becomes energetically prohibitive in the vicinity of the constraining  substrate. Cavities may  thus break away from the substrate and  continue to grow into the bulk, where they first become optically visible. This mechanism may also help to explain the consistent vicinity of experimentally observed cavitation events to the bead interface.
Though  subtle, this distinction between nucleation in the bulk or at an interface has significant implications for interpreting the mechanisms underlying material failure in a wide range of reinforced composites, and serves to  shift the focus from bulk to interfacial properties in optimizing material performance. }
Finally, as stated by \cite{gent1990cavitation}, \textit{``the fact that a theory appears to work does not mean that it is true”}. In this spirit, future work is needed to be able to further elucidate the sub-micron phenomena that are currently below the optical limit, and to resolve the potential involvement of fracture in the bulk as an additional  failure mechanism.

\section*{Acknowledgments}
\noindent We acknowledge the support of our work through the Office of Naval Research under grant No. N000142312530, the National Science Foundation under Award No. CMMI1942016 and The Army Research Office under grant No. W911NF2310089.

\appendix
\setcounter{figure}{0} 
\section{Spherical cap formula} \label{app:sphere}
\noindent To obtain the relationship $S_O=S_O(V,d)$ based on the spherical cap approximation of the shape of the expanded interfacial cavities, we first write  
\begin{equation}\label{eqA1:1}
    S_O=\pi h^2, \qquad \text{and}\qquad V=\frac{\pi}{6}h\left(\frac{3}{4}d^2+h^2\right),
\end{equation}
where $h$ denotes the height of the cap, and recall that $S_O$ is the change of surface area. The second of the above formula can be recast in the inverted (yet less pleasant) form
\begin{equation}
    {h}=\frac{d}{2}\left({(g(V/d^3))^{1/3}-(g(V/d^3))^{-1/3}}\right), \qquad \text{where} \qquad g(x)=\frac{24}{\pi} x +\sqrt{1+\left(\frac{24}{\pi} x\right)^2}.
\end{equation} Here we have define the auxiliary function $g$. 
Upon substitution of the above result in the first equation of \eqref{eqA1:1} we obtain the following analytical formula for $S_o$
\begin{equation}\label{eqA1:SO}
    S_O= \frac{\pi d^2}{4}\left({(g(V/d^3))^{1/3}-(g(V/d^3))^{-1/3}}\right)^2.
\end{equation}

\end{document}